# The International Linear Collider Machine Staging Report 2017

Addendum to the International Linear Collider Technical Design Report published in 2013

Linear Collider Collaboration / October, 2017
Editors:Lyn Evans and Shinichiro Michizono

# Table of Contents

# 1. Introduction

The Technical Design Report (TDR) of the ILC mainly concentrates on a baseline machine of 500 GeV centre-of-mass with detailed cost and manpower estimates consistent with this option. However, the discovery of a Higgs Boson with a mass of 125 GeV opens up the possibility of reducing cost by starting at a centre-of-mass energy of 250 GeV with the possibility of future upgrades to 500 GeV or even 1 TeV+. The rest of this paper outlines the options for the design of a 250 GeV "Higgs factory". The scientific programme that the machine would offer is the subject of a separate report[1].

A first stage 250 GeV machine would imply the installation of approximately half of the linac of the 500 GeV baseline machine. There are several possible scenarios for the civil construction and conventional facilities of which three are considered in this report.

Option A
Only the tunnel for the 250 GeV machine is constructed and equipped. Increasing the machine energy by extending the tunnel length would then require extensive additional civil engineering at a later date.

Option B
The tunnel length is extended to allow the energy to be increased to 350 GeV (the top quark threshold) at a later date by adding more acceleration structure. Only the downstream part is filled with linac. The remaining tunnel will be left in a bare state (no dividing wall, cooling or ventilation) in order to save money in the first stage. Upgrading the energy to 350 GeV then requires finishing the tunnel and installing extra cavities.

Option C
The complete tunnel and access shafts for the 500 GeV machine is constructed in the beginning and only the downstream part is filled with linac. The remaining tunnel will be left in a bare state (no dividing wall, cooling or ventilation), the same as Option B, in order to save money in the first stage. Upgrading the energy to 500 GeV then requires finishing the tunnel and installing extra cavities.

The first scenario (Option A) represents the lowest cost for the initial phase. The second and third obviously require extra investment in the initial stage but open up a simple possibility of increasing the centre-of-mass energy without major tunneling work.

The main parameters, including luminosity, are initially assumed to be the same as those specified for the 500 GeV baseline scaled to 250 GeV (Table 12.1 of the TDR). This means that the electron and positron sources, damping rings and bunch compressors remain unchanged from the baseline. However, an improved luminosity performance has been worked out and is described in Section 5 of this report.

The beam delivery systems could, in principle, be further optimized for low energy but the overall geometry is still assumed to be consistent with an eventual 1 TeV upgrade.

For positron production, a 5 Hz is still assumed but the lower energy of the electron beam (125 GeV instead of 150 GeV in the baseline) makes it more difficult to produce the required polarized positron flux. To compensate for the lower energy, the undulator length must be increased by about 60% in order to preserve the photon flux to the convertor target. A more straightforward way of preserving the positron flux would be to use a conventional positron source which would require an additional 3 GeV linac. This option would mean that partial polarization of the positrons would not be possible. The impact on the scientific potential of the machine must be addressed. First indications are that the costs of the two options are very similar.

This addendum to the TDR published in 2013 provides a brief summary of the ILC staging study focusing on the staging energy of 250 GeV with varieties of energy extendability up to 500 GeV.

## 2. Positron production options

### *2.1 Undulator: baseline design*

The TDR baseline design produces positrons by transporting the primary electron beam through a superconducting helical undulator. The overall layout of the positron source is shown schematically shown in Figure 2-1. In this TDR configuration we expect the minimal electron energy to be 150 GeV and positrons with 30% polarization to be generated.

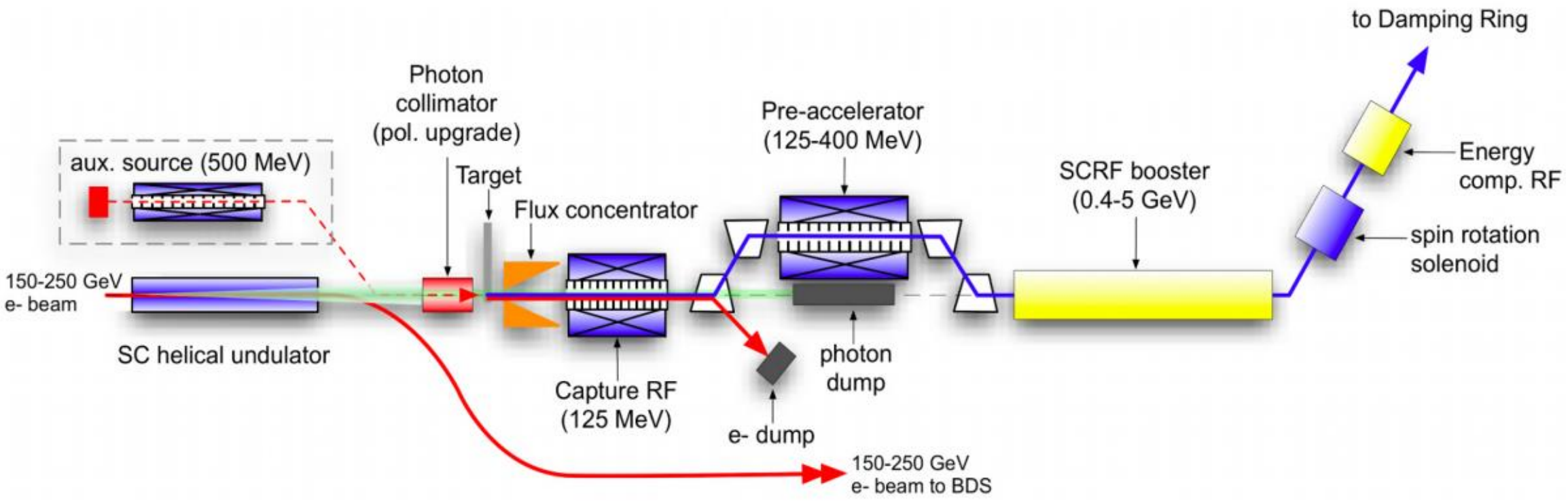

Figure 2-1 Overall layout of the undulator source of positrons.

In the case of a 125 GeV electron beam, the required positron flux can be generated by increasing the undulator length from 147 m to 231 m. This longer undulator positron source is the new baseline for the ILC250GeV staging. Electrons lose ~3 GeV in the undulator and this is compensated by the main electron linac. Owing to this, the number of RF units in the main electron linac is higher than that in the positron linac (already in TDR). The collision timing constraint (described later) should be satisfied in this undulator scheme.

### *2.2 Conventional: alternative design*

An electron-driven (e-driven; conventional source of positrons) design is an alternative to the TDR undulator concept for positron production. Although polarized positrons are not available in this scheme, positron beam commissioning is possible without a 125 GeV electron source.

The e-driven source of positrons consists of a normal conducting (NC) 3 GeV linac, a positron target, and a normal conducting 5 GeV linac. The energy of the driving linac was re-designed to be suitable for operation of 1,312 bunches, leading to a reduction in the linac energy from 4.8 GeV to 3 GeV[2]. The system is shown schematically in Figure 2-2.

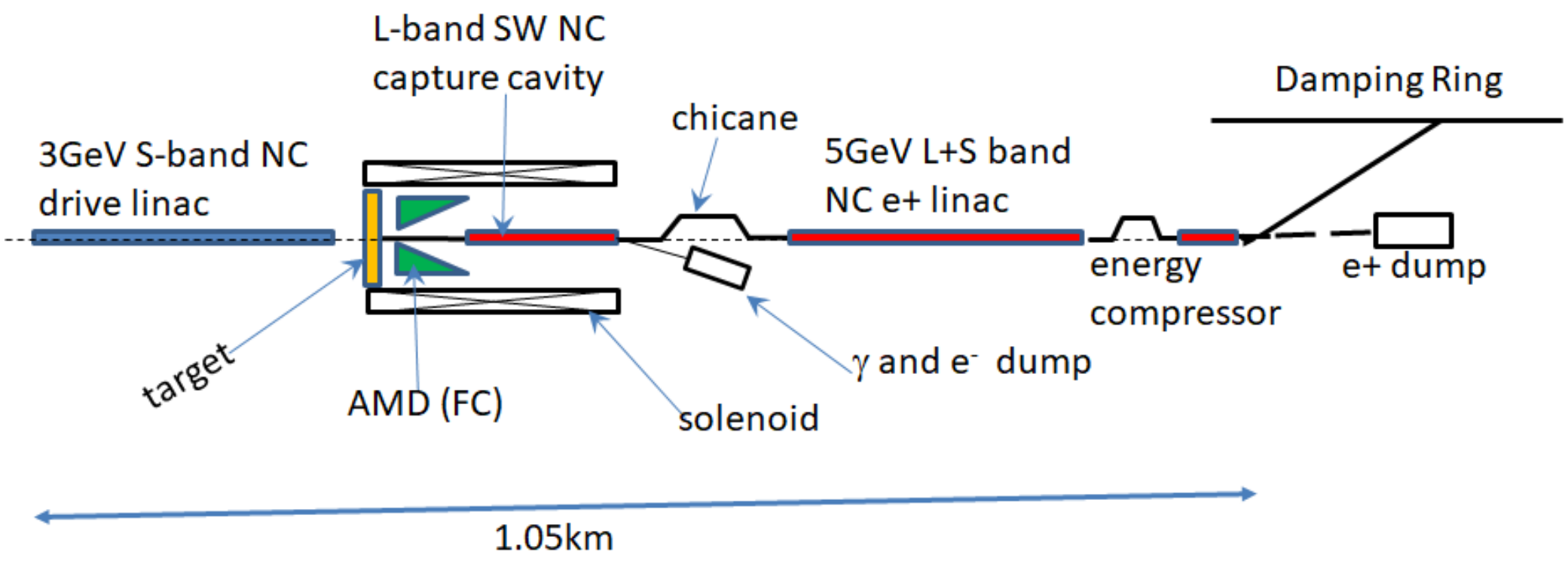

Figure 2-2 Layout of the e-Driven source of positrons.

In this e-driven scheme, different electron bunch patterns will be used, as shown in Figure 2-3. Beam pulses with ~480 ns duration (including ~66 bunches) will be accelerated in the normal conducting linacs. The linacs will operate at 20 pulses every 200 ms, with inter-pulse intervals of 3.3 ms. The remaining 137 ms will be reserved for damping of positrons in the damping ring.

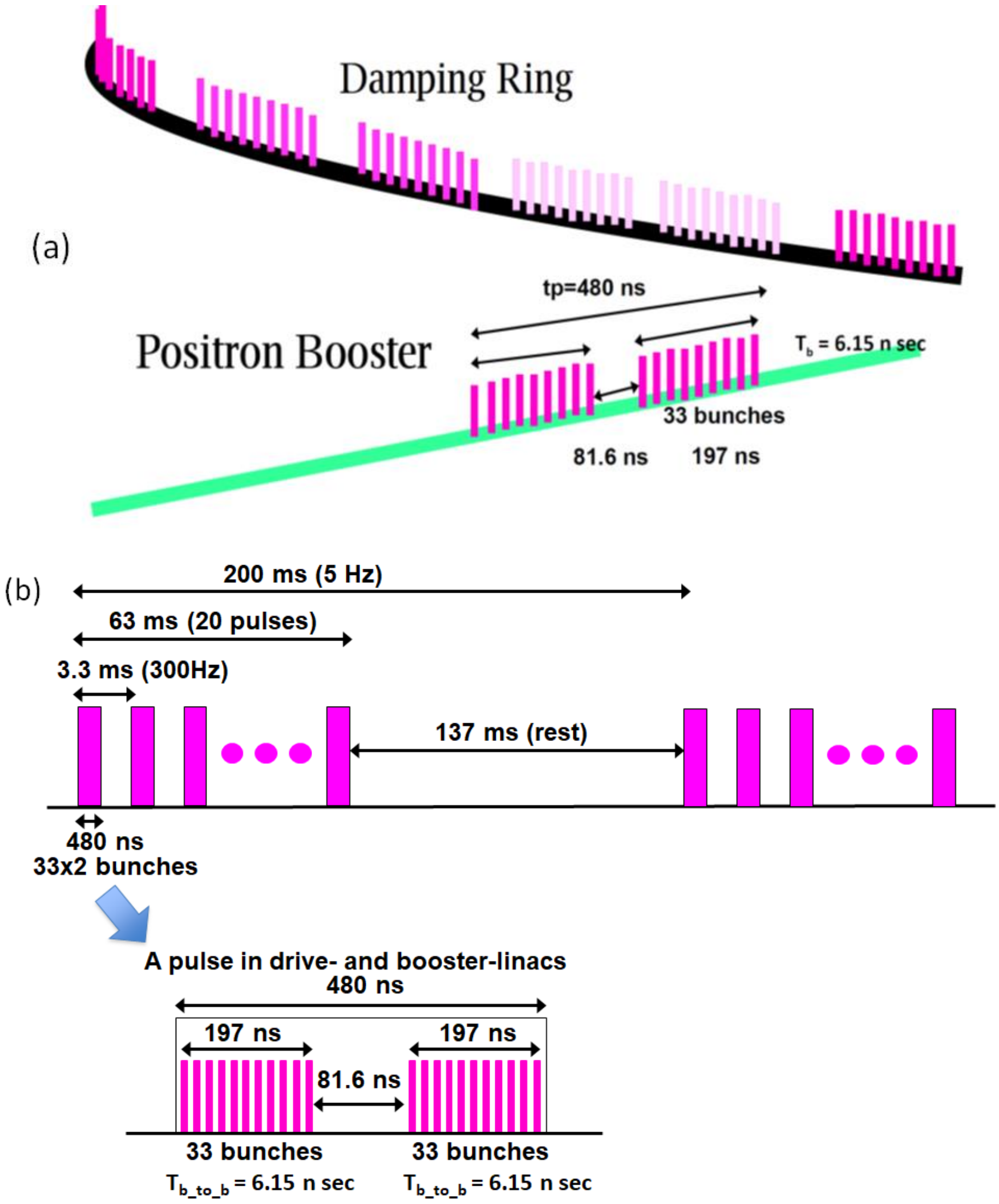

Fig.2-3 Schematic of the beam injection bunches (a) and beam bunch structure (b).

## *2.3 Luminosity upgrade schemes*

Following several years of successful operation of the initial ILC250 a luminosity upgrade is possible. The basic change in the luminosity upgrade is the increase in the number of bunches from 1,312 to 2,625. Since the margin of the electron cloud instability for the positron damping ring is ~3, the number of bunches can probably be doubled without the second positron damping ring in the undulator scheme. We will obtain sufficient information on the necessity of the second ring from superKEKB and first stage ILC.

In the case of the e-driven source of positrons, one more positron damping ring is required because beam-loading compensation is difficult to realize with a 3-ns-wide bunch spacing. In addition, the driving beam linac should be extended from 3 GeV to 4.8 GeV and the modulators of the driving linac and booster should be reinforced owing to longer beam pulse durations.

## *2.4 Cost comparison*

We found that there is no cost difference between accelerator components for the undulator and e-driven source of positrons. Some cost reduction (of the order of a few ten's of MILCU*) associated with the e-driven system is expected, if the space for the timing constraint in the undulator scheme is omitted.

The undulator source will still be considered as the baseline source of positrons. However, an e-driven source of positrons could be adopted initially for ILC250 GeV and be replaced by the undulator source in future upgrades, depending on the technical maturity, because the e-driven source is safer for achieving design luminosity at low electron energies (~125 GeV) and has the big advantage that positron beam commissioning can be done without needing the full electron linac and damping ring. However, it has the disadvantage of no positron polarization.

*The reference currency (the "ILCU") is the United States dollar (USD) as of January, 2012.

# 3 Variants of the baseline (Options A/B/C)

## *3.1 Accelerator configuration*

The accelerator configuration is shown schematically in Figure 3-1. The change requests post-TDR are included in the baseline design (TDR update).

- A reduced ML tunnel cross-section is adopted and the central shield wall is changed from 3.5 m to 1.5 m.
- A vertical shaft access to the detector hall is adopted.
- A collision timing constraint (required for the undulator source of positrons) is satisfied.

A TDR-undulator-based positron source is used. This has a collision timing constraint. The length of the undulator is changed from 147 m to 231 m to produce positrons using a 125-GeV-energy beam.

Only the operation of a 5 Hz linac (not a 10 Hz one as envisaged in the TDR) is considered, for maximal cost reduction. The maximal individual cryoline length is 2.5 km ± 10%, the same as for TDR. The non-staging areas are kept untouched (i.e., the $e^-$ source, DR, turn-around, bunch compressor, BDS, and IR).

Option A is a minimal configuration for the ILC250GeV. Option B has a 350-GeV-energy tunnel, and the accelerators are located downstream. A simple tunnel is extended upstream in Option B. Normal wall finish, air-conditioning, lighting, and water drainage will be installed but the central shield wall, AC power line, and cooling water line will not be installed. Option C has a 500-GeV-energy tunnel and accelerators are located at the downstream side.

The average accelerating gradient 31.5MV/m is assumed for each of these options as in TDR. The cases where 35MV/m is assumed after successful R&D are named Option A', B', and C'.

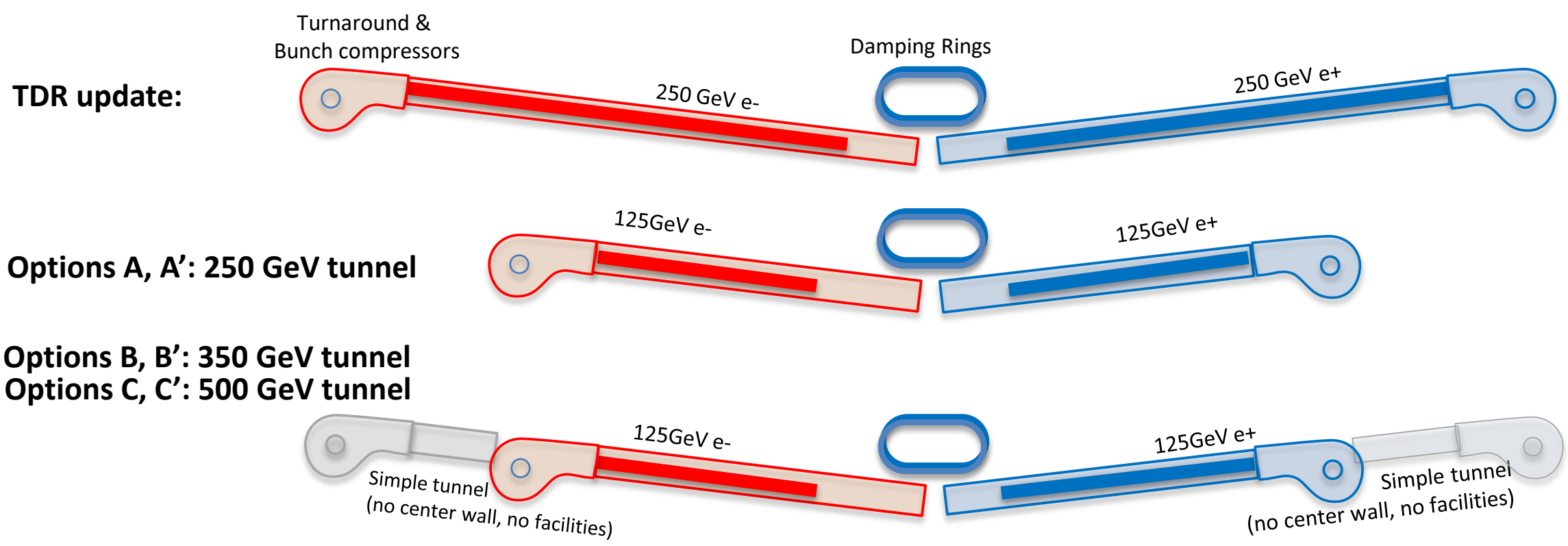

Figure 3-1 Staging options.

## *3.2 Collision timing constraint*

To collide $e^-/e^+$ at the IP, the collision timing constraint in the case of the undulator $e^+$ source has to be satisfied. This constraint is schematically shown in Figure 3-2. The following relationship should be satisfied:

$$(L_1+L_2+L_3)-L_4=n \text{x} C_{DR}$$

We assume that the damping ring circumference remains unchanged ($C_{DR}$ = 3,238.68 m),though there is still a possibility to change it.

The TDR (after the change request) has n = 10. In this case, 1,473 meters of space for the $e^+$ ML (and the $e^-$ ML for the energy symmetric upgrade in the future) are added for this adjustment. In the case of Option A, n = 6 is adopted. There is an adjustment space of 583 m in each ML. In the case of Option B, n = 6 (for the energy of 250 GeV) and n = 8 (for the energy of 350 GeV) are adopted. This corresponds to an additional simple tunnel of 3,238 m in each ML. In the case of Option C, n = 6 (for the energy of 250 GeV) and n = 10 (for the energy of 500 GeV) are adopted. This corresponds to an additional simple tunnel of 6,477 m in each ML.

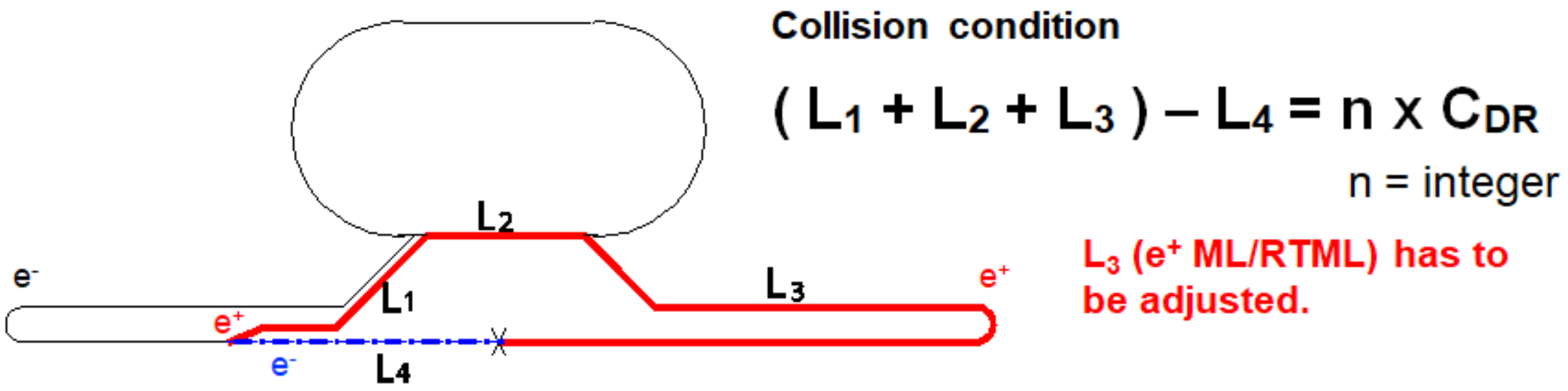

Figure 3-2 Collision timing constraint.

## 3.3 Energy margin

The energy reach margin for sufficient generation of positrons and safe energy reach of Higgs physics at 250 GeV must be included.

(1) Module margin: We introduced a 2.5% margin (corresponding to the energy of 3.1 GeV in each linac) to reach the target energy of the target experiment (no margin was included in the TDR).

(2) Availability margin: Three RF units were expected to compensate for a cryomodule trip in the TDR (1.5%). The same three RF units (117 Cavities, 13.5 cryomodules) are counted as the availability margin corresponding to 3% in ILC250GeV.

(3) Space margin: This is the cryomodule space to allow for more cryomodules to be installed in the future. The vacant space (to satisfy the collision timing constraint) can be filled with cryomodules in later stages (by introducing additional cryogenics and RF systems)

At all times, 0.5% is required to compensate for cavity phase offset. Therefore, the total energy balance is 2% (=1.5%+0.5%) for ILC500GeV and 6% (=2.5%+3%+0.5%) for ILC250GeV.

## 3.4 Effect of cost-reducing R&D

An average cavity gradient of 31.5 MV/m is assumed in the TDR. The staged ILC250 design considers both the cavity gradient of 31.5 MV/m (TDR) and 35 MV/m (in the case of successful cost-reducing R&D). Nevertheless, the overall tunnel length is the same, owing to the collision timing constraint (n = 6). The space of 1,049 m in each ML in the 35 MV/m configuration will be used for the "space margin". We will not change the RF configuration system even in the case of the 35 MV/m scheme. These baseline configurations are summarized in Table 3-1. Figures 3-3~3-7 show the configurations of the TDR and Options A, B, C, A'. The configurations of options B' and C' are the combinations of Option A' and the simple tunnel in Options B and C.

Table 3-1: Summary of baseline configurations.

| Options | Gradient [MV/m] | $E_{CM}$ [GeV] | Total $E_{CM}$ Margin | n | Space margin | Reserved tunnel (each end) | Total tunnel |
|---|---|---|---|---|---|---|---|
| TDR update | 31.5 | 500 | 2% | 10 | 1,473 m | 0 m | 33.5 km |
| Option A | | 250 | 6% | 6 | 583 m | 0 m | 20.5 km |
| Option B | | | | 6&8 | | 3,238 m | 27 km |
| Option C | | | | 6&10 | | 6,477 m | 33.5 km |
| Option A’ | 35 | | | 6 | 1,049 m | 0 m | 20.5 km |
| Option B’ | | | | 6&8 | | 3,238 m | 27 km |
| Option C’ | | | | 6&10 | | 6,477 m | 33.5 km |

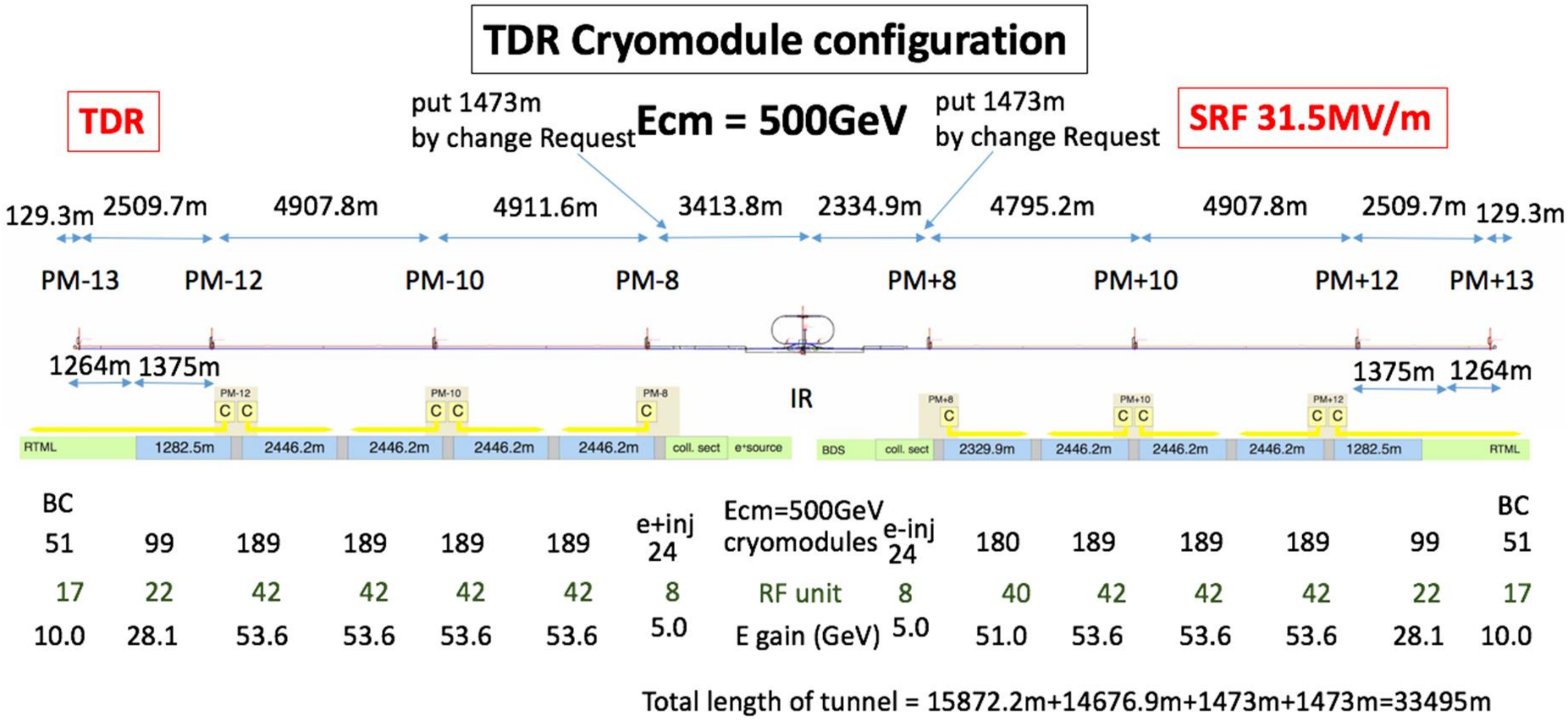

(note: 1 ML unit = 2 RF unit = 9 CMs (in TDR) )
Figure 3-3 TDR (ILC 500GeV) configuration.

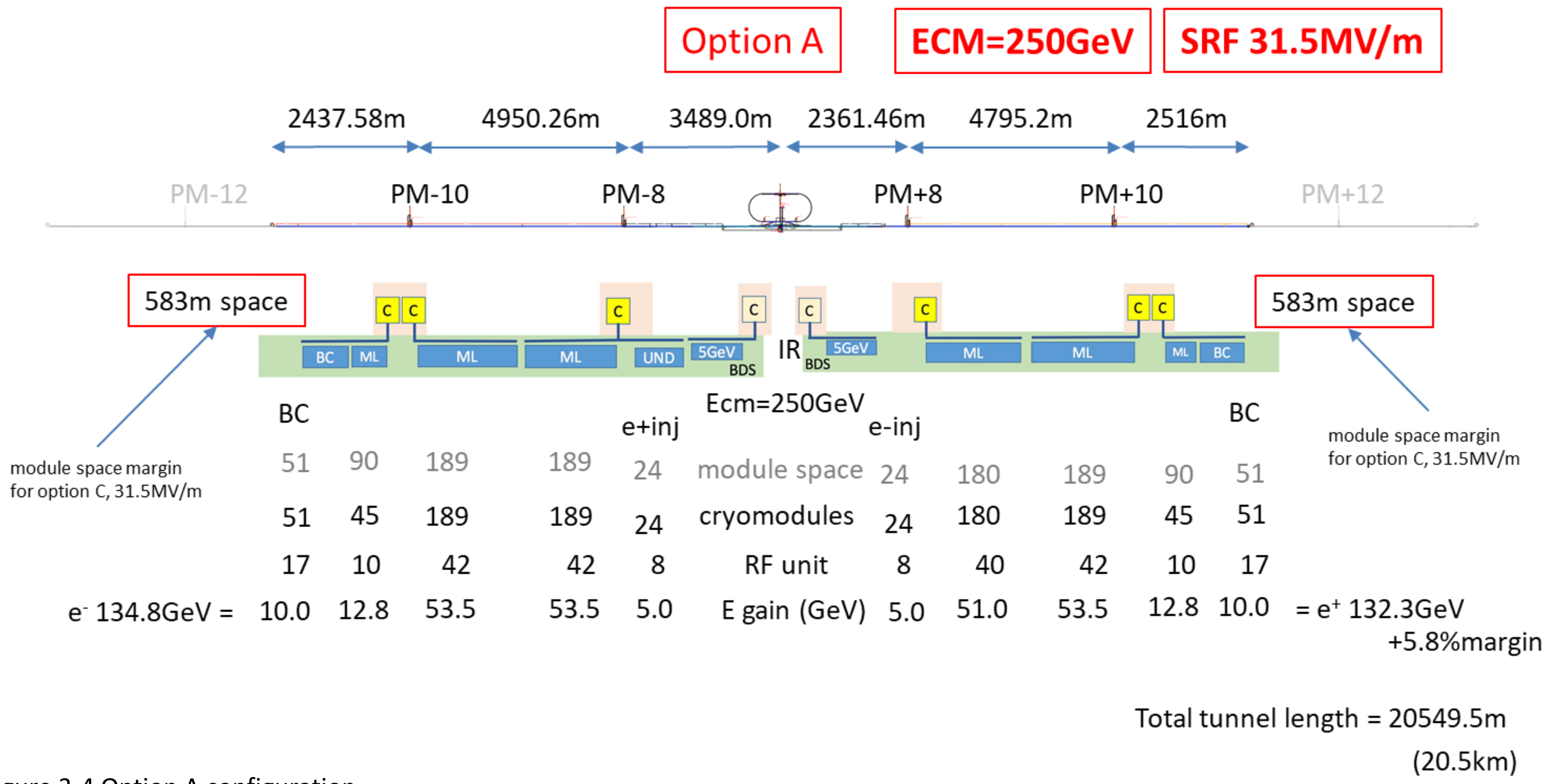

Figure 3-4 Option A configuration.

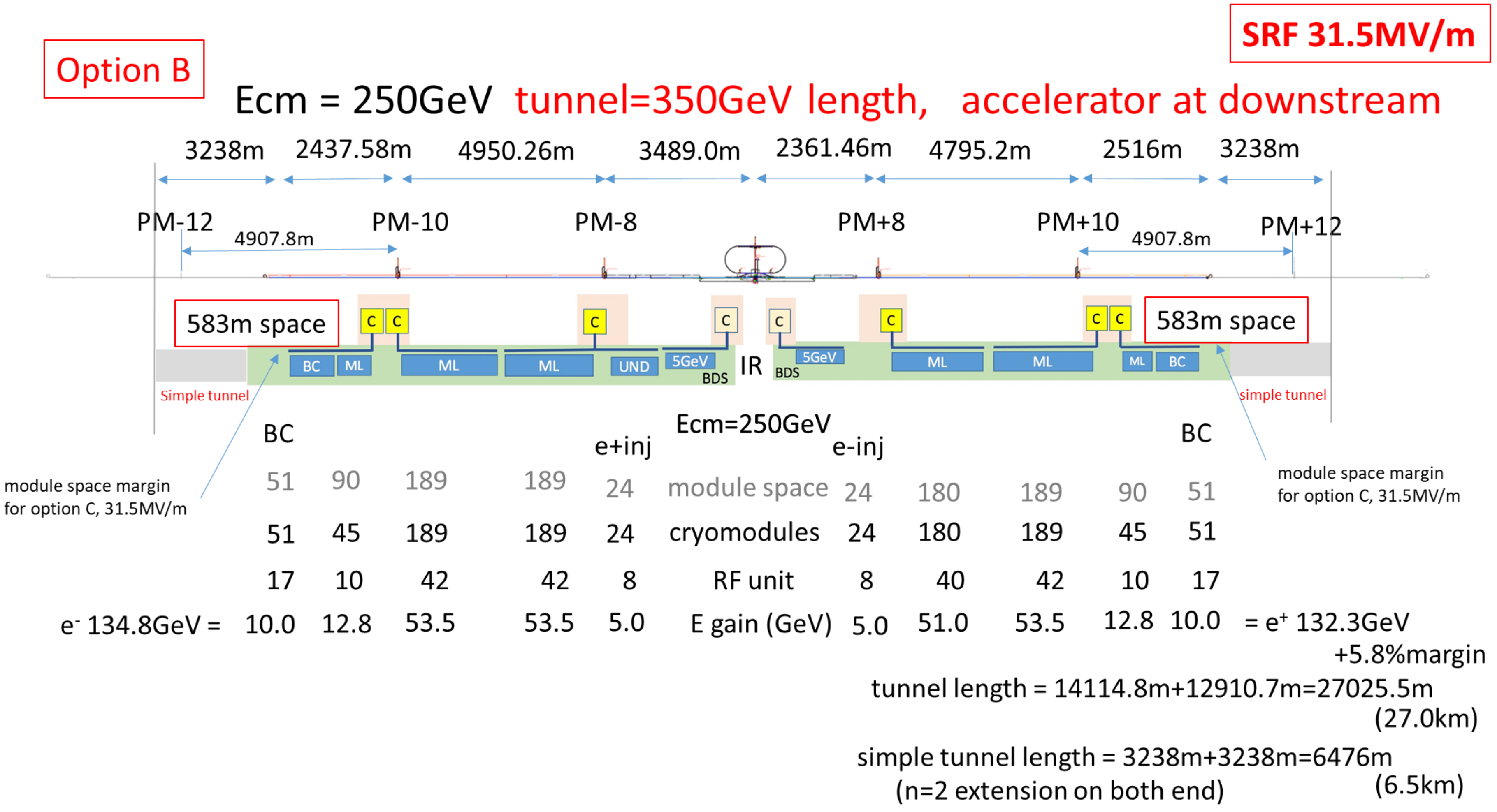

Figure 3-5 Option B configuration.

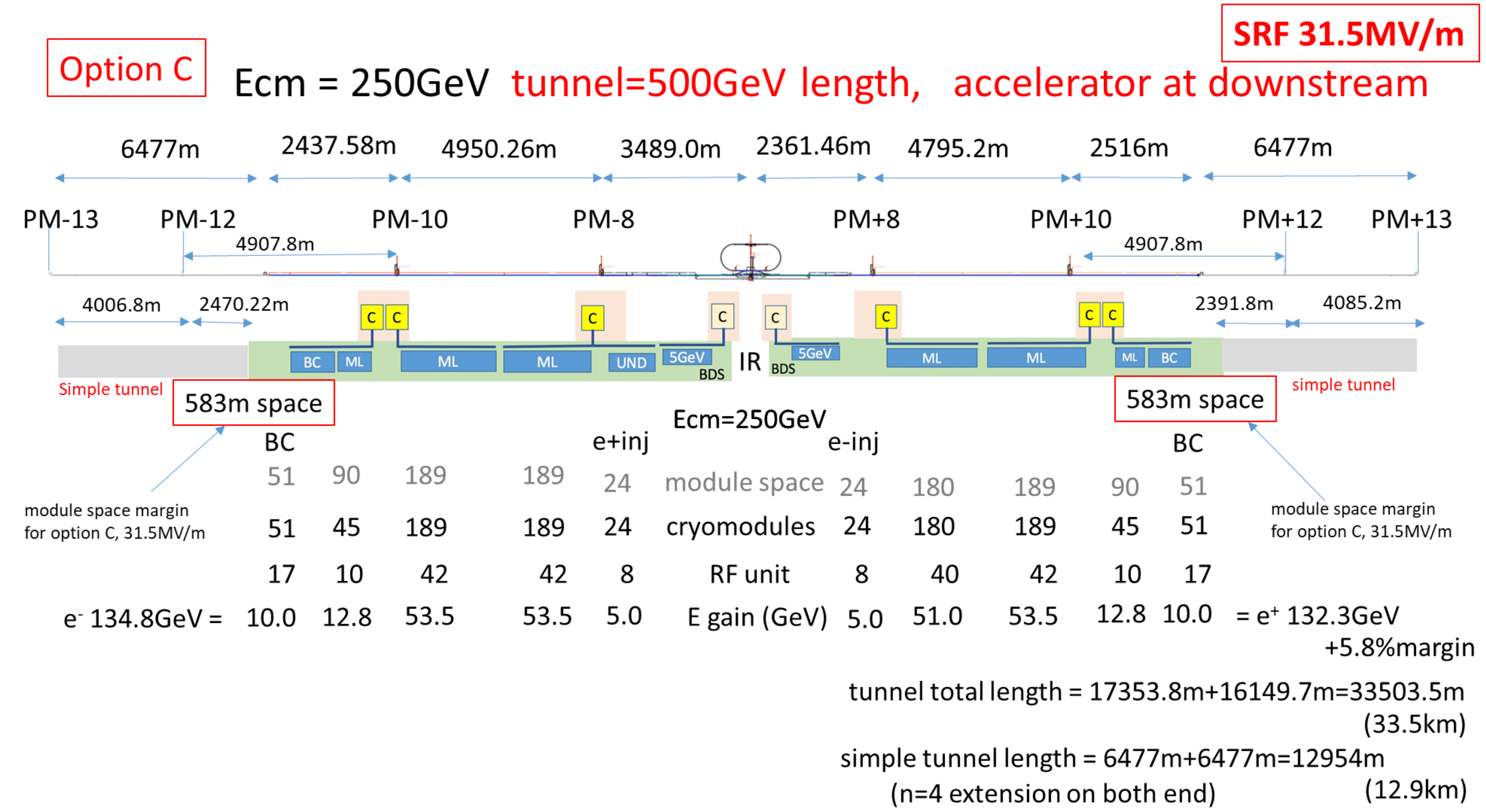

Figure 3-6 Option C configuration.

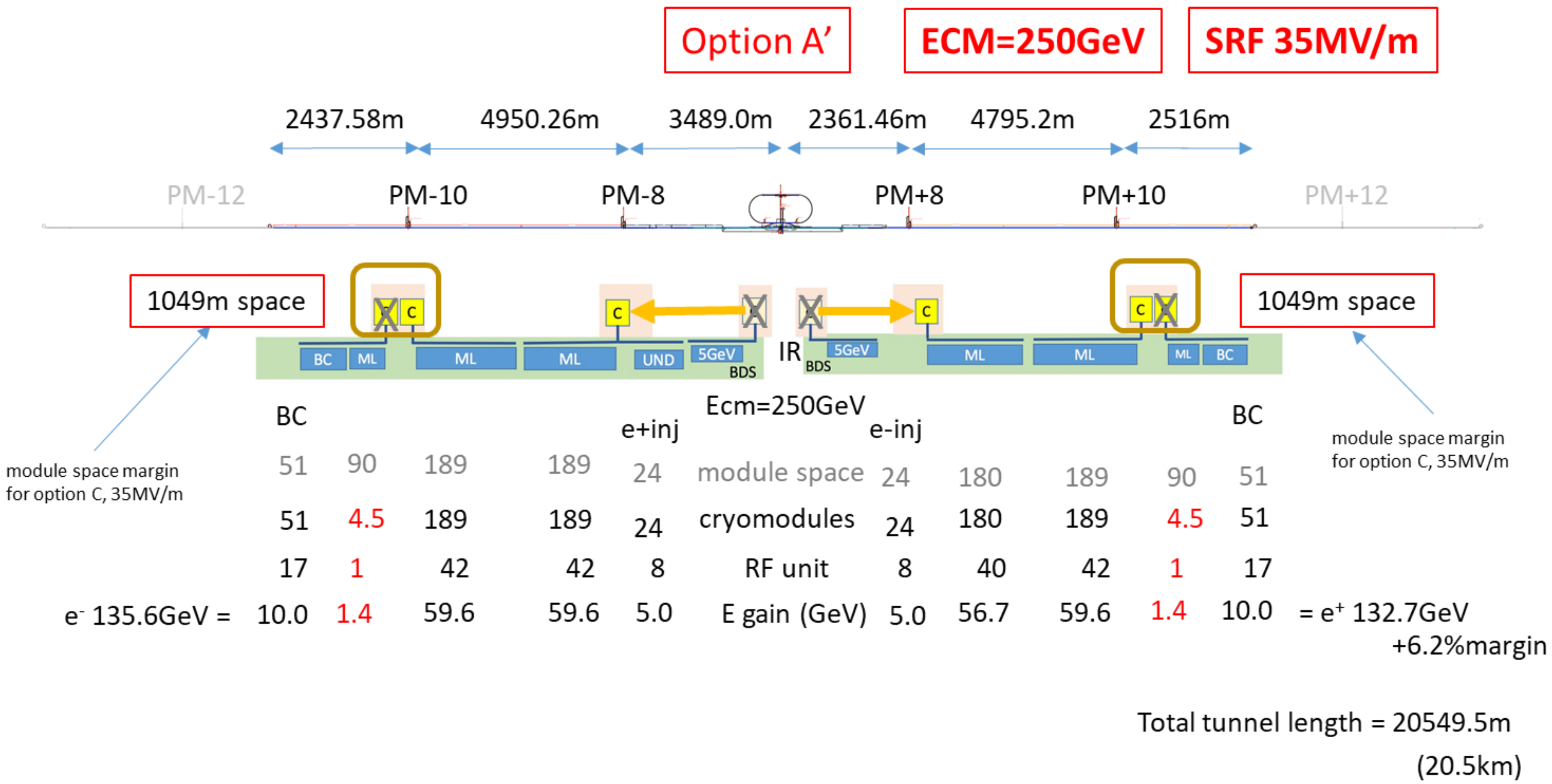

Figure 3-7 Option A’ configuration.

## 4. SRF R&D and resulting cost reduction

The main fraction of the accelerator construction cost is attributed to the main linac (ML) with superconducting RF (SRF) technology. Thus, our main focus for cost reduction is associated with the SRF technology. We consider the following four areas in R&D:

### *4.1 Preparation of niobium materials (processing for sheet fabrication and piping)[3]*

The cost of niobium materials for fabrication of SRF cavity cells and end groups is relatively high, owing to the use of a rare material and an elaborated preparation process. R&D aims to reduce the cost of materials by optimizing the production of Nb ingots and by optimizing the disk/sheet and pipe forming process, to prepare for cavity fabrication. The TDR (and also XFEL, LCLS-II) specified the residual resistivity ratio (RRR) to be >300. Low RRR material limits the maximum cavity gradients, and we propose to optimize the purity of ingots with a lower RRR (> 200 and 250 on average) with acceptable specific residual content such as Ta, and to simplify the fabrication of Nb sheets/disk using direct slicing from Nb ingots, to maintain clean surfaces, by eliminating forging, rolling, and mechanical polishing/grinding processes. We expect a major cost reduction related to the fabrication of Nb sheets for SRF cavities, as shown in Figures 4-1 and 4-2.

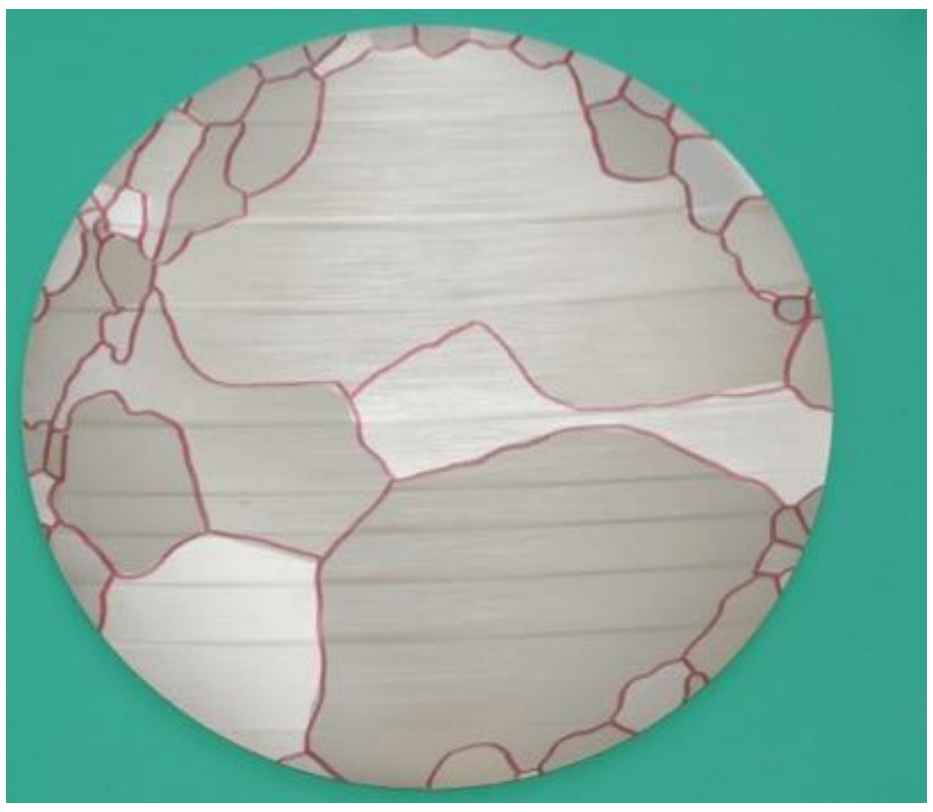

Figure 4-1 Large-grain Nb sheet.

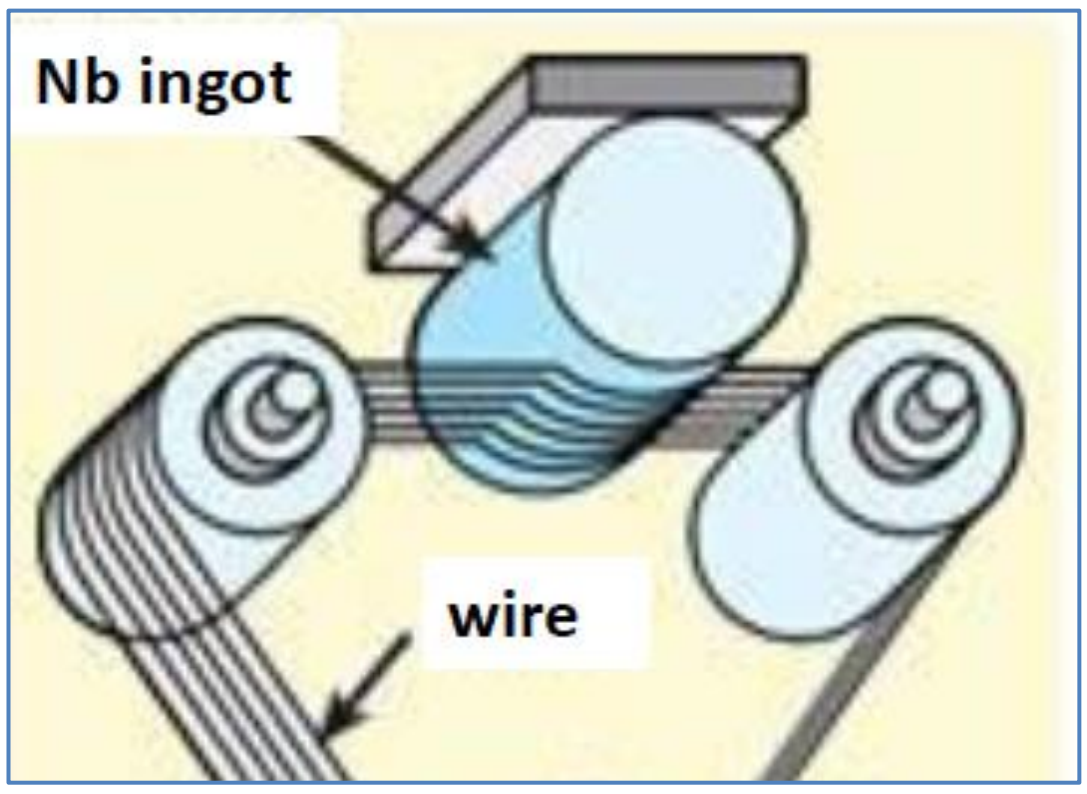

Figure 4-2 Direct slicing from the Nb ingot.

### *4.2 SRF cavity fabrication to ensure a high gradient and high Q (with a new surface process demonstrated by Fermilab)[4]*

Recent SRF cavity R&D results reported by Fermilab have shown that improvements in the accelerating gradients by 10% and above, and Q by a factor of 2, with respect to standard ILC cavity treatment may be possible. Performance characteristics of cavities are shown in Figure 4-3. Improvement is achieved by modified 120°C vacuum baking of cavities with nitrogen infusion at 120 °C, directly after a heat treatment process at ~ 800°C. The new treatment should make it possible to operate cavities at higher gradients, thus reducing the SRF linac length. Because the new process can eliminate the second round of electro-polishing (EP), reduction in the chemical surface treatment cost is also expected. Higher Q and a flatter Q versus $E_{acc}$ curve could lower the cost of cryogenics and ensure more cost-effective operation.

We expect an average 35 MV/m cavity gradient (up from TDR's 31.5 MV/m at the TDR), where $Q_0 \sim 1.6 \times 10^{10}$ ($0.8 \times 10^{10}$ at 35 MV/m in the TDR) leading to a smaller number of cavities and cryogenic systems.

The same RF distribution system (shown in Figure 4-4) can be used in the 35 MV/m operation. However, we are planning to develop a higher efficiency klystron (with a maximal power of 11 MW and efficiency of 71%). The development of such high-efficiency klystrons is anticipated at world-wide accelerator laboratories, and offers potential energy consumption savings.

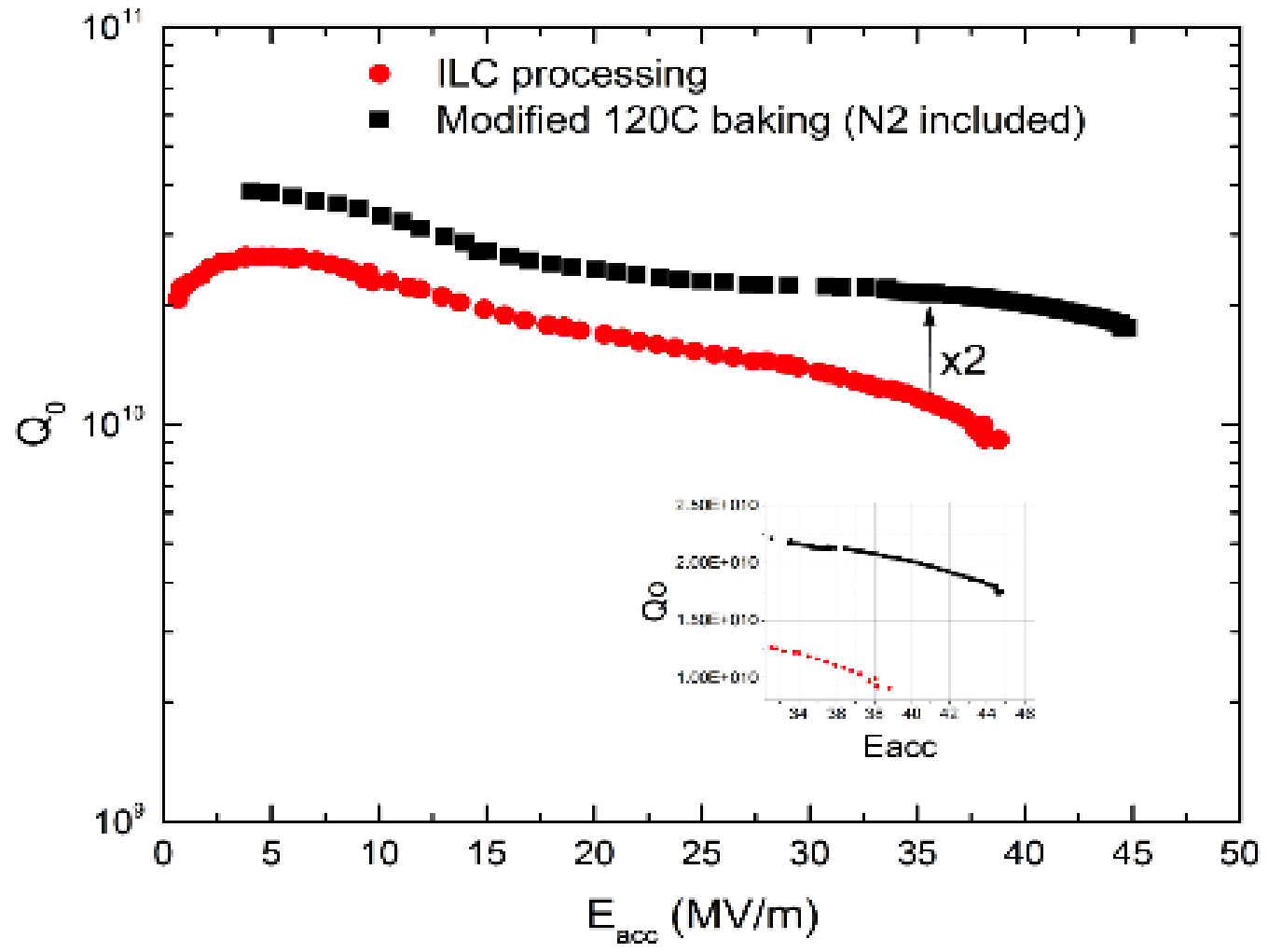

Figure 4-3 Higher performance with N-infusion treatment.

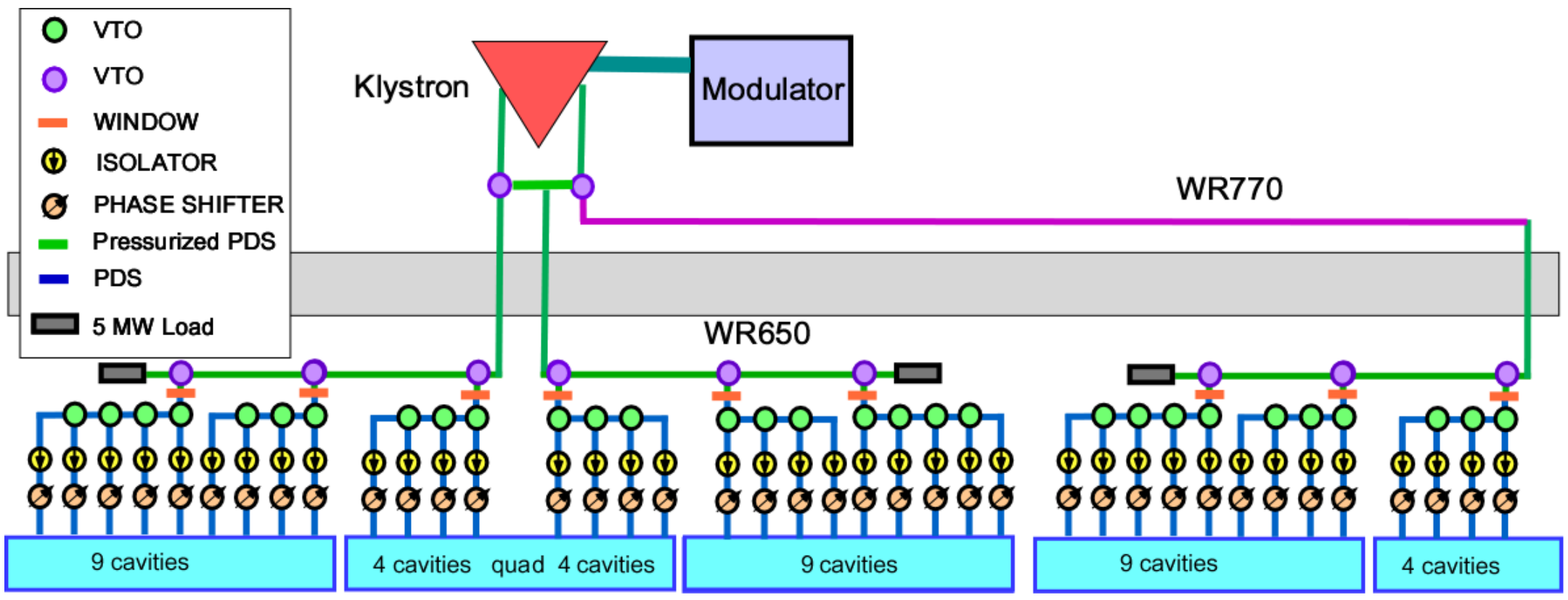

Figure 4-4 RF distribution system at the main linac.

## 4.3 Power input coupler fabrication[5]

R&D aims to optimize the material (for the ceramic windows) and the Cu-plating procedure, as well as the mechanical design for cost effective assembly with the SRF cavity string. New ceramic window material, with low secondary electron emission yield, developed in Japan, is promising for the fabrication cost reduction, because the coupler can be fabricated without an anti-multipactor coating (TiN).

## 4.4 Cavity chemical treatment[6,7]

The change in the SC-cavity chemical treatment from using Horizontal electropolishing (EP) and Sulphur-acid+HF as proposed in the TDR, to Vertical EP + non-HF solution + Bipolar EP, will lead to substantial process cost reduction. This involves development of cathode electrodes and non-HF with a bipolar voltage power supply for smooth surfaces. The effects are: simpler infrastructure, shorter processing time, cheaper solution, cheaper solution waste process, and a safer process without HF.

If the R&D will be successful, we expect a ~10% reduction compared with the TDR accelerator cost at ILC500GeV. In the case of ILC250GeV, the reduction would be ~6% (compared with the ILC500 TDR cost), because the number of SRF systems at the main linac would be halved and the cryogenics design would be further optimized.

## 5. Improvement of Luminosity

The accelerator described in the TDR is based on the best optimization for $E_{CM}$ = 500 GeV. The machine parameters at lower energies are given in Table 2.1 (vol.3.II) but they are basically obtained by scaling from 500 GeV. Some improvement of the machine design parameters may be achieved through re-optimization for $E_{CM}$ = 250 GeV for the first stage.

We consider re-optimization of luminosity at 250 GeV. The luminosity is proportional to $P_B$/E x $(\delta_{BS}$ / $\varepsilon_{ny})$ $^{1/2}$, where $P_B$ is the beam power, $\delta_{BS}$ the loss of energy associated with beamstrahlung, and $\varepsilon_{ny}$ the normalized vertical emittance. To increase $P_B$ is costly and to decrease $\varepsilon_{ny}$ requires tighter alignment tolerance of the main linac. Therefore, we choose to accept a larger $\delta_{BS}$, which is still small (~1%) at 250 GeV. The best way to increase $\delta_{BS}$ is to reduce the horizontal beam size at the IP by reducing the horizontal normalized emittance $\varepsilon_{nx}$ (on the other hand, reducing the horizontal beam size by reducing $\beta_x$ would cause synchrotron radiation background). This is achieved by slightly changing the design of the damping rings (using longer dipole magnets in the arcs). This will not cause a significant cost increase (B*L, magnetic field times length, is the same).

The new set of parameters is listed in Table 5-1 together with the TDR sets of parameters for 250 GeV and 500 GeV. The resulting luminosity is 1.35 × $10^{34}$ /$cm^2$/s at 250 GeV, a factor 1.65 higher than that for the TDR (note: the values for the TDR have been corrected by the Change Request 5).

Several issues are under study:

- The vertical disruption parameter $D_y$ is now ~35, higher than ~25 in the TDR. This will require a more accurate beam position feedback at the IP.
- The background owing to the larger $\delta_{BS}$ and the increase in the number of incoherent pairs may significantly affect the detector's performance.

The change of $\varepsilon_{nx}$ will not increase the luminosity at 500 GeV, because $\delta_{BS}$ is already large (4.5%), but tuning the final focusing system will become somewhat easier because $\beta_x$ can be relaxed.

Table 5-1: New beam parameters optimized for ILC250GeV.

| | | | TDR | | New |
|---|---|---|---|---|---|
| Center-of-mass energy | $E_{CM}$ | GeV | 250 | 500 | 250 |
| Bunch population | N | e10 | 2 | 2 | 2 |
| Bunch separation | | ns | 554 | 554 | 554 |
| Beam current | | mA | 5.78 | 5.78 | 5.78 |
| Number of bunches per pulse | Nb | | 1312 | 1312 | 1312 |
| Collision frequency | | Hz | 5 | 5 | 5 |
| Electron linac rep rate | | Hz | 10 | 5 | 5 |
| Beam power (2 beams) | $P_B$ | MW | 5.26 | 10.5 | 5.26 |
| r.m.s. bunch length at IP | $\sigma_z$ | mm | 0.3 | 0.3 | 0.3 |
| relative energy spread at IP (e−) | $\sigma_E/E$ | % | 0.188 | 0.124 | 0.188 |
| relative energy spread at IP (e+) | $\sigma_E/E$ | % | 0.15 | 0.07 | 0.15 |
| Normalized horizontal emittance at IP | $\varepsilon_{nx}$ | μm | 10 | 10 | 5 |
| Normalized vertical emittance at IP | $\varepsilon_{ny}$ | nm | 35 | 35 | 35 |
| Beam polarization (e−) | | % | 80 | 80 | 80 |
| Beam polarization (e+) | | % | 30 | 30 | 30 |
| Beta function at IP (x) | $\beta_x$ | mm | 13 | 11 | 13 |
| Beta function at IP (y) | $\beta_y$ | mm | 0.41 | 0.48 | 0.41 |
| r.m.s. beam size at IP (x) | $\sigma_x$ | nm | 729 | 474 | 516 |
| r.m.s. beam size at IP (y) | $\sigma_y$ | nm | 7.66 | 5.86 | 7.66 |
| r.m.s. beam angle spread at IP (x) | $\theta_x$ | μr | 56.1 | 43.1 | 39.7 |
| r.m.s. beam angle spread at IP (y) | $\theta_y$ | μr | 18.7 | 12.2 | 18.7 |
| Disruption parameter (x) | Dx | | 0.26 | 0.26 | 0.51 |
| Disruption parameter (y) | Dy | | 24.5 | 24.6 | 34.5 |
| Upsilon (average) | $\Upsilon$ | | 0.020 | 0.062 | 0.028 |
| Number of beamstrahlung photons | $n_\gamma$ | | 1.21 | 1.82 | 1.91 |
| Energy loss by beamstrahlung | $\delta_{BS}$ | % | 0.97 | 4.50 | 2.62 |
| Geometric luminosity | Lgeo | $e34/cm^2s$ | 0.374 | 0.751 | 0.529 |
| Luminosity | L | $e34/cm^2s$ | 0.82 | 1.79 | 1.35 |

# 6. Cost estimate for ILC 250 GeV

## *6.1 The cost of accelerator construction*

The cost estimate is carried out with the ILCU. The TDR defined the "ILCU" as the United States dollar (USD) as of January, 2012. RF unit cost and other unit cost is calculated from TDR. The staging cost is obtained by subtracting the decreased number of units. Reduced volume production effect and price fluctuation from 2012 are ignored because these depend on the different components.

The construction cost for Option A is lower by ~34% compared with ILC 500GeV. In this estimate, 6% of total energy margin from having additional cryomodules is included (2% for the ILC 500 GeV case) for sufficient generation of positrons and safe energy reach for Higgs physics at 250 GeV, as mentioned in Section 3.3. In addition, the "space margin" is reserved owing to the timing constraint shown in Table 3.1. If we add the cost reduction in the SRF system resulting from the R&D effort, the expected cost reduction for Option A' becomes ~40%. It should be noted that the length of the tunnel is kept the same as Option A.

As for the human resources, Option A (A') requires 75% of those needed for the TDR 500 GeV baseline.

For Options B (B') and C (C'), the expected construction cost reductions are ~33% (~39%) and ~31.5% (~37.5%), respectively, with a similar level of reduction in human resources requirement to that obtained for Option A.

Table 6-1: Summary of the staging cost

| | e+/e- collision [GeV] | Tunnel Space for [GeV] | Value Total (MILCU) | Reduction [%] |
|---|---|---|---|---|
| TDR | 250/250 | 500 | 7,980 | 0 |
| TDR update | 250/250 | 500 | 7,950 | -0.4 |
| Option A | 125/125 | 250 | 5,260 | -34 |
| Option B | 125/125 | 350 | 5,350 | -33 |
| Option C | 125/125 | 500 | 5,470 | -31.5 |
| Option A' | 125/125 | 250 | 4,780 | -40 |
| Option B' | 125/125 | 350 | 4,870 | -39 |
| Option C' | 125/125 | 500 | 4,990 | -37.5 |

The Value estimates broken down by the following system, i.e. Main Linac (ML), the electron and positron Ring to Main Linac (RTML), Common, Damping Ring (DR), Beam Delivery System (BDS), Positron source, Electron source and Interaction Region (IR), are shown in Figure 6-1. The cost reduction from the TDR is mainly coming from the main linac owing to the smaller SRF system and shorter tunnel length. The difference from Option A to A' (B to B', and C to C') results from the cost reduction R&D. Simple and empty tunnels are added to the upstream side in the case of Options B and C (B' and C'), resulting in the cost difference between Options A, B, and C. "Common" consists of common parts in the ILC laboratory, such as the main campus, the main AC power station, general computing system (laboratory networking, e-mail system, business computers etc.), accelerator installation and control systems. The main campus and computing system costs are saved according to the reduction in human resources. The installation and control system costs are saved due to the reduction of the main linac energy. Some slight cost reduction at the electron and positron sources between A and A’ (or B and B’, C and C’) results from the SRF system used in these sources.

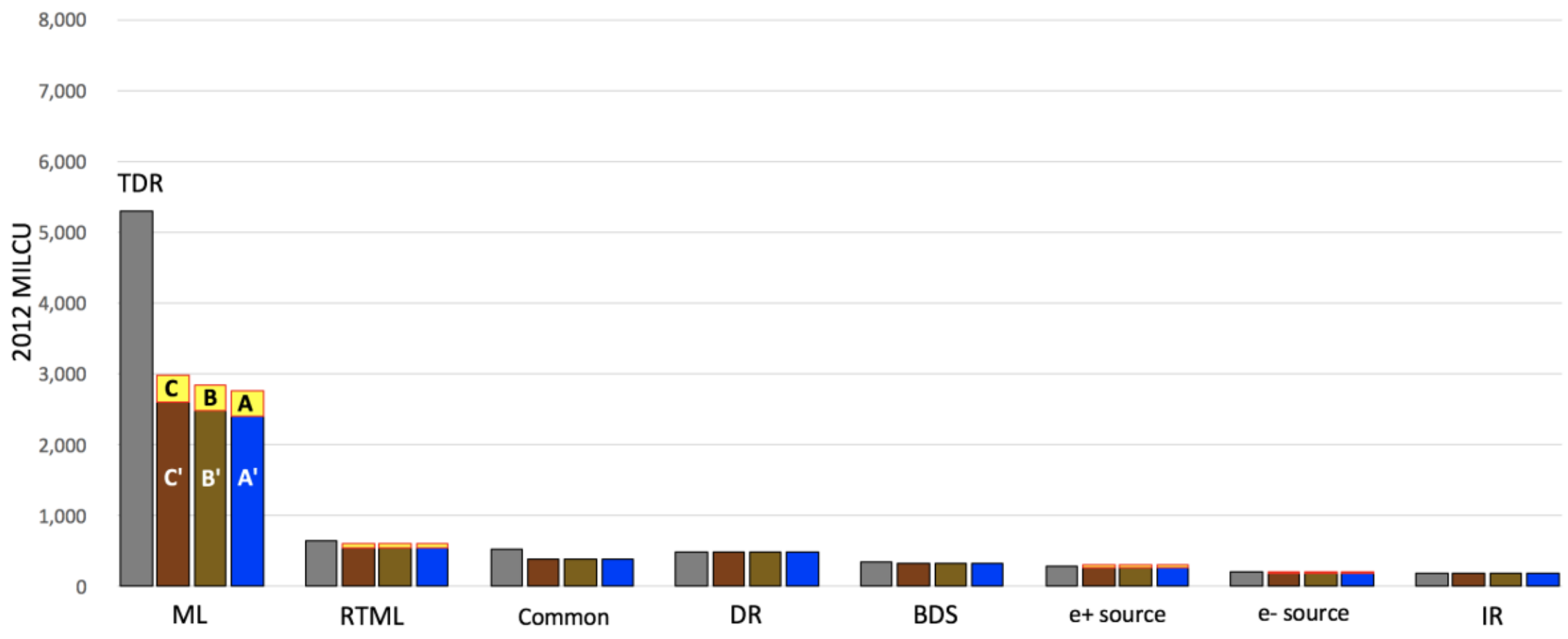

Figure 6-1 Distribution of the ILC value estimate by area system. The numbers give the estimate for each system in MILCU.

## *6.2 Operational cost*

Electric power for the ILC 500 GeV operation was ~164 MW. The electric power in Options A and A’ will be ~129 MW and ~ 125 MW or less, respectively. The operational cost may then be reduced by ~23%, and by even more than 25% if the SRF cost-reduction R&D will be successful.